\documentclass[runningheads]{llncs}
\usepackage[T1]{fontenc}
\usepackage{graphicx}
\usepackage{marvosym}
\usepackage{amsfonts}
\usepackage{amsmath}
\usepackage{booktabs}
\usepackage{makecell}
\begin{document}
\title{Cross-Modal Transformer GAN: A Brain Structure-Function Deep Fusing Framework for Alzheimer's Disease}

\author{Junren Pan, Shuqiang Wang}

\authorrunning{Anonymous}

\institute{Shenzhen Institutes of Advanced Technology, Chinese Academy of Sciences, Shenzhen 518000, China
\\
\email{sq.wang@siat.ac.cn}}
\titlerunning{CT-GAN}
\maketitle              
\begin{abstract}
Cross-modal fusion of different types of neuroimaging data has shown great promise for predicting the progression of Alzheimer's Disease(AD).
However, most existing methods applied in neuroimaging can not efficiently fuse the functional and structural information from multi-modal neuroimages.
In this work, a novel cross-modal transformer generative adversarial network(CT-GAN) is proposed to fuse functional information contained in resting-state functional magnetic resonance imaging (rs-fMRI) and structural information contained in Diffusion Tensor Imaging (DTI).
The developed bi-attention mechanism can match functional information to structural information efficiently and maximize the capability of extracting complementary information from rs-fMRI and DTI.
By capturing the deep complementary information between structural features and functional features, the proposed CT-GAN can detect the AD-related brain connectivity, which could be used as a bio-marker of AD.
Experimental results show that the proposed model can not only improve classification performance but also detect the AD-related brain connectivity
effectively.

\keywords{Cross-modal fusion \and Transformer \and Bi-attention Mechanism \and Brain Network \and Generative Adversarial Strategy }
\end{abstract}
\section{Introduction}

Alzheimer's disease (AD), a chronic and irreversible neurodegenerative disease, is the main reason for dementia among aged people~\cite{DadarPascoal2017}.
According to statistical data given by Alzheimer's Association\cite{AS}, at least 50 million people worldwide are suffering from AD.
AD patients will gradually lose cognitive function such as remembering or thinking, and eventually become unable to take care of themselves~\cite{Association2018}.
The widespread incidence of AD makes a severe financial burden to both patients' families and governments.
With the development of artificial intelligence~\cite{ml1,ml2,ml3,ml4,ml5,ml6}, researchers study AD from different angles using machine learning technology~\cite{mci1,mci2,mci3}.
However, the cause of AD has not been fully revealed.
One of the main reasons for the above difficulties is that brain is a highly complex network, and completing cognitions requires specific coordination between regions-of-interest (ROIs).

A brain network can be characterized as a graph.
The nodes of a brain network represent ROIs of the brain.
The edges of a brain network represent the interaction relationship between ROIs of the brain.
There are two basic connectivity categories of brain networks: functional connectivity (FC) and structural connectivity (SC).
FC is defined as the interdependence between the blood-oxygen-level-dependent (BOLD) signals of two ROIs, where BOLD signals can be extracted from rs-fMRI.
SC is defined as the neural fibers connection strength among ROIs, which can be extracted from DTI.
Many studies \cite{sh3,JeonKang2020,sh2} have used FC or SC to obtain some AD-related features that can not be discovered in traditional imaging methods.
This shows that brain network methods have more advantages than the traditional imaging method in AD research.
However, most existing brain network studies are based on a single modal,
which can only focus on one of the brain structural information and brain functional information.
Since single modality data may only contain complementary cross-modal information partially,
it will lose an opportunity to take advantage of complementary cross-modal information to study AD more accurately.
Therefore multimodal brain network methods~\cite{ZhangShen2012,multi1,multi2,multi3} are gaining more and more attention in medical imaging computing.
For the structure-function deep fusing task of multimodal brain networks, the key is how to efficiently use complementary cross-modal information that is heterogeneous and hidden in different types of neuroimaging data.
Most existing structure-function fusion approaches just used linear relationships between different modalities.
However, changes of brain structure and function can not be fully characterized by linear relationships.
Previous studies\cite{Honey,Li15} proved that strong SC inclines to be accompanied with strong FC, but not vice versa.
On the other hand, clinical studies\cite{Da2015,LeiCheng2020,Cao2018} show that when an SC between RIOs is reduced, some regions can increase functional activity to compensate for the reduced SC.

To overcome the above problem, a novel cross-modal transformer is proposed in this work to deal with the structure-function deep fusion task based on generative adversarial networks(GANs).
GANs can bee seen as variational-inference\cite{bftd1,bftd2} based  generative model. 
GANs are proved to be an efficient framework for learning complex distribution \cite{GAN,shgan1}.
Currently, GANs are successfully used in many branches of medical image analysis~\cite{shgan2,shgan3,shgan4,shgan5,shgan6,shgan7}.
Transformers~\cite{trans1} have shown their powerful capability for sequential analyzing in natural language processing (NLP).
This is due to the self-attention mechanism which characterizes the nonlinear relationships between given inputs.
Following their successful applications in the area of NLP, transformers have been adopted to image tasks very recently~\cite{trans2,trans3,trans4}.
However, transformers have been few explored in the area of brain networks.
In this study, a cross-modal transformer is proposed to fuse structure-function information of brain networks.
The proposed cross-modal transformer consists of the following four modules:
1)convolutional neural networks(CNN) modules that are used to extract functional information from rs-fMRI;
2)graph convolutional networks(GCN) modules that are used to extract structural information from DTI;
3) F2S-attention modules that transform from functional information to structural information;
4) S2F-attention modules that transform from structural information to functional information.
The cross-modal transformer is based on a bi-attention mechanism where complementary information from rs-fMRI and DTI are fused layer by layer.
On the other hand, a generative adversarial strategy is adopted to guide the proposed model's training.
The advantages of the proposed model are summarized as follows:
1)structural and functional information can be deeply fused;
2)complementary information from rs-fMRI and DTI can be effectively extracted;
3)due to the generative adversarial strategy, the proposed model does not need very deep nets, which make the model's architecture and training more flexible.

\section{Method}
\subsection{The Architecture}
\begin{figure}
\includegraphics[width=\textwidth]{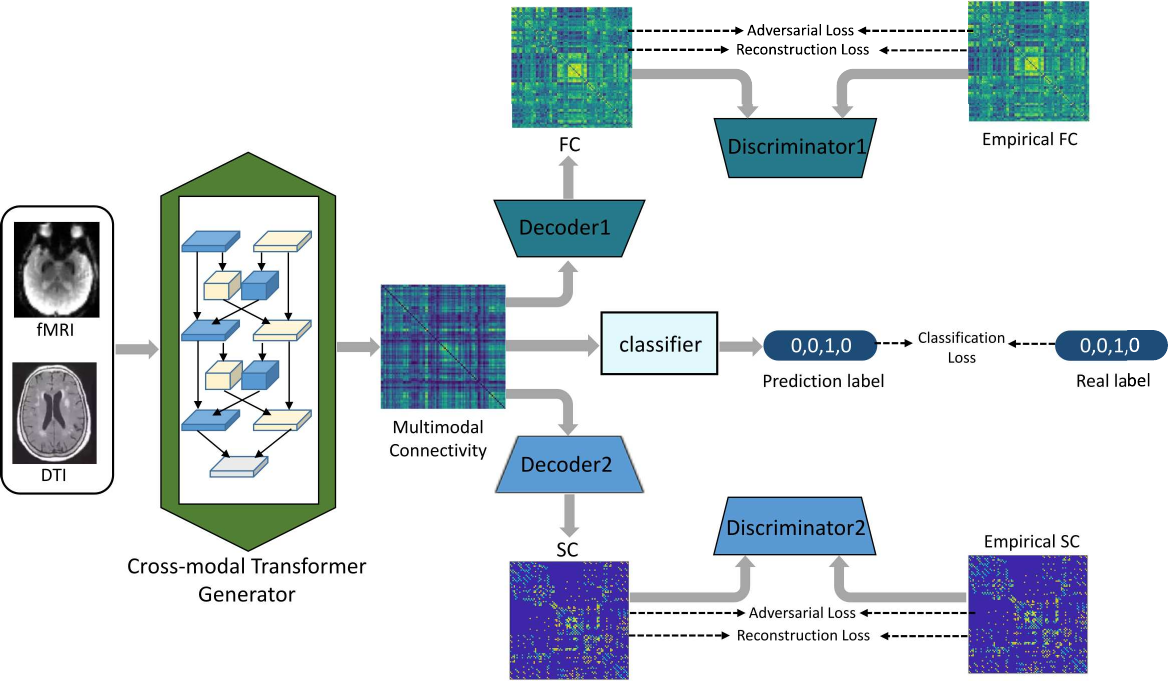}
\caption{The framework of the proposed CT-GAN.
} \label{fig1}
\end{figure}

The proposed CT-GAN is illustrated in Fig.~\ref{fig1}, which consists of four components:
1) a cross-modal transformer generator that outputs multimodal connectivity brain networks;
2) two decoders that decode multimodal connectivity to the corresponding SC and FC, respectively;
3) two discriminators, one of which determines whether an SC from learned by our proposed model or output by a software template, the other discriminator for FC is similar;
4) a classifier that predicts AD stages according to multimodal connectivity.

\subsection{Bi-Attention Mechanism}
The key idea of this work is to exploit the bi-attention mechanism of transformers to fuse structure-function information for rs-fMRI and DTI given their complementary nature.
A transformer mapping an input feature sequence as $X=\mathbb{R}^{n\times d_X}$ to a target feature sequence as $Y=\mathbb{R}^{n\times d_Y}$,
where $n$ is the total number of ROIs,
can be described as follows.
Firstly, a linear projections is used to compute a set of queries matrix $Q$, keys matrix $K$, and values matrix $V$,
\begin{equation}
Q = XW^q,\quad K = XW^k,\quad V = XW^v
\end{equation}
where $W^q \in \mathbb{R}^{d_X\times d_q}$ , $W^k \in \mathbb{R}^{d_X\times d_k}$ with $d_k = d_q$, and $W^v \in \mathbb{R}^{d_X\times d_v }$ are weight matrices.
The attention of $Q$, $K$, and $V$ can be obtained as follows:
\begin{equation}
\text{Attention}(Q,K,V) = \text{softmax}\Big(\frac{QK^T}{\sqrt{d_k}}\Big)V.
\end{equation}
And then, a fully connected layer(FC) is used to transform the attention of $Q$, $K$, and $V$ into a feature sequence
with the same dimension as the target feature sequence $Y$.
Finally, the output of a transformer is
\begin{equation}
X^{out} = \text{FC}\big(\text{Attention}(XW^q,XW^k,XW^v)\big) + \lambda Y,
\end{equation}
where $\lambda$ is a hyper-parameter between 0 to 1.

\begin{figure}[h!]
\includegraphics[width=\textwidth]{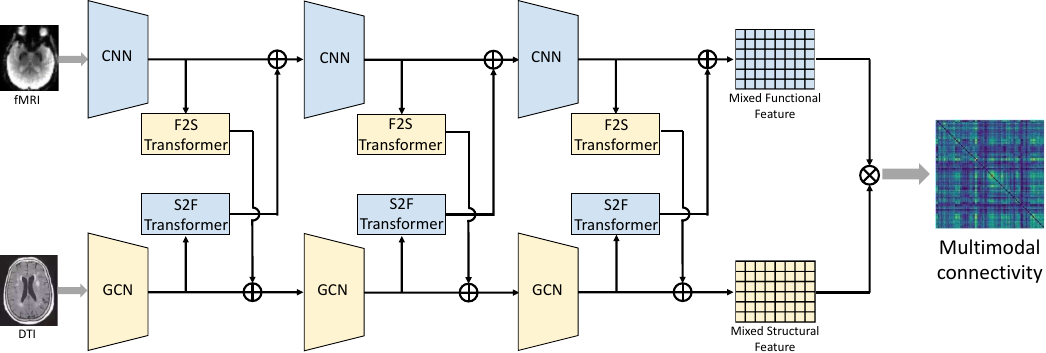}
\caption{The network architecture of proposed generator.
} \label{gene}
\end{figure}

There are two types of transformers in the proposed cross-modal generator. The one is used to transform functional information into structural information,  abbreviated as F2S.
The other is used to transform structural information into functional information, abbreviated as S2F.
We can now introduce the bi-attention mechanism of transformers.
BOLD signals extracted from rs-fMRI and empirical structural connectivity extracted from DTI are inputs to our cross-modal generator that uses several pairs of F2S and S2F modules for fusing intermediate features between the inputs.
The intermediate features of BOLD signals and empirical structural connectivity are extracted by CNN and GCN, respectively.
The detailed architecture of generator is shown in Fig.~\ref{gene}.
Let us denote the mixed functional feature sequence and mixed structural feature sequence in Fig.~\ref{gene} to be $F$ and $S$, respectively.
It is worth mentioning that $F\in \mathbb{R}^{n\times d}$ represents the feature sequence of each ROI and $S\in \mathbb{R}^{n\times n}$ represents the connective feature between ROIs where $n$ is the total number of ROIs.
The multimodal connectivity matrix MC as the final output of the proposed generator is obtained by
\begin{equation}
\text{MC} = SFF^TS^T.
\end{equation}

\subsection{Loss Function}
Success in structure-function deep fusion tasks requires semantic reasoning.
Therefore a multimodal connectivity matrix learned by the proposed model should be able to decouple to be the corresponding empirical FC matrix and empirical SC matrix.
A generative adversarial strategy is used to achieve the above process.
To guarantee the quality of generated multimodal connectivity,
a hybrid loss function is used to optimize the network, including three types of terms:
the adversarial loss, the classification loss, and the pair-wise connectivity reconstruction loss.
The details are given as follows.

\textbf{Adversarial Loss.} To make the FC matrix and SC matrix decoded by multimodal connectivity matrix as close as possible to empirical FC matrices and SC matrices,
the adversarial losses are defined as
\begin{equation}
\mathcal{L}_{adv}^{\text{SC}} = \mathbb{E}_{x}[\log D_1(\text{SC}_x)]+\mathbb{E}_{x}[\log (1- D_1(Dec_1(G(x))))],
\end{equation}
\begin{equation}
\mathcal{L}_{adv}^{\text{FC}} = \mathbb{E}_{x}[\log D_2(\text{FC}_x)]+\mathbb{E}_{x}[\log (1- D_2(Dec_2(G(x))))],
\end{equation}
where $G$ is the generator, $D_1$ and $D_2$ are the discriminator1 and discriminator2, $Dec_1$ and $Dec_2$ are the decode1 and decode2 in Fig.~\ref{fig1}.
$SC_x$ and $FC_x$ represent the empirical SC matrix and empirical FC matrix of subject $x$.
The generator $G$ generates a multimodal connectivity matrix $G(x)$ from subject $x$'s rs-fMRI and DTI.
The decoders $Dec_1$ and $Dec_2$ decode $G(x)$ into an FC matrix and an SC matrix, respectively.
While the discriminator $D_1$ attempts to distinguish between the FC matrix decoded by $Dec_1$ and empirical FC matrices($D_2$ similarly works on SC matrices).
The generative adversarial strategy is that $G$ tries to minimize above adversarial losses while $D$ tries to maximize it.

\textbf{Classification Loss.}For multimodal connectivity matrices, an important index to judge the effect of cross-modal fusing is whether they can achieve high accuracy in predicting AD stages.
The classification loss is imposed when optimizing the classifier and the generator $G$.
The formula of classification loss is given by
\begin{equation}
\mathcal{L}_{cls} = \mathbb{E}_{(x,y)\sim p_{\text{real}}(x,y)}[-\log p_c(y|G(x))],
\end{equation}
where $y$ represents AD stages, including normal controls(NC), early mild cognitive impairment(EMCI), late mild cognitive impairment(LMCI), and Alzheimer's Disease(AD).
The $p_c(y|G(x))$, output by the classifier with the input $G(x)$, represents the probability that the subject $x$ is now under stage $y$.
By classification loss, the generator is trained to extract and fuse features, which contains more disease information, from rs-fMRI and DTI.
Meanwhile, the classifier can achieve the highest predicting accuracy through multimodal connectivity matrices.

\textbf{Pair-wise Connectivity Reconstruction Loss.}The $L^1$ pair-wise connectivity reconstruction loss is used to impose an additional topological constraint on the generator $G$.
It means that the generator $G$ and the decoders $Dec1, Dec2$ are not only needed to fool the discriminators $D1$ and $D2$.
In addition, they also need to minimize the sum of pair-wise connectivity difference between FC/SC matrices decoded by $Dec1/Dec2$ and empirical FC/SC matrices.
The $L^1$ pair-wise reconstruction losses are formalized as follows:
\begin{equation}
\mathcal{L}_{rcs}^{\text{FC}} = \mathbb{E}_{x}\|\text{FC}_x - Dec_1(G(x))\|_1,
\end{equation}
\begin{equation}
\quad \mathcal{L}_{rcs}^{\text{SC}} = \mathbb{E}_{x}\|\text{SC}_x - Dec_2(G(x))\|_1.
\end{equation}

\section{Experiments}
DTI and rs-fMRI used to validate the proposed framework are from the public dataset ADNI(Alzheimer's Disease Neuroimaging Initiative).
There are 268 subjects' data we used in this study whose detailed information can be seen in Table~\ref{information}.
AAL 90 atlas is used in the preprocessing.
A string of preprocessing steps of rs-fMRI using the DPARSF toolbox is consisted of discarding the first 20 volumes, head motion correction, band-pass filtering, Gaussian smoothing, and extracting time series of all voxels.
The structural connectivity is obtained by tracking fiber bundles between ROIs.
The fiber tracking stopping conditions set in PANDA, the toolbox used to preprocess DTI, are following that:
(1) the crossing angle between two consecutive moving directions is more than 45 degrees.
(2) the fractional anisotropy value is not in the range of [0.2, 1.0].
\begin{table}[h]
\caption{Subjects' information in this study}

\label{information}

\centering
\begin{tabular}{c c c c c c}

\toprule[2pt]

  \textbf{Group} & \textbf{AD}(63) & \textbf{LMCI}(41) & \textbf{EMCI}(80) & \textbf{NC}(84) \\

\midrule[1pt]

  Male/Female & 39M/24F & 20M/21F & 48M/32F & 38M/46F   \\

  Age(mean $\pm$ SD)  & 75.3 $\pm$ 5.5 & 74.9 $\pm$ 5.3 & 75.8 $\pm$ 6.1 & 74.0 $\pm$ 5.9   \\

\bottomrule[2pt]

\end{tabular}

\end{table}

The decoders $Dec_1$ and $Dec_2$, the discriminators $D_1$ and $D_2$, the classifier are implemented by multilayer perceptron.
The generator $G$ consists of three blocks. Such blocks contain a CNN layer with kernel size (1,10), a GCN layer, an F2S transformer, and an S2F transformer.
The hyper-parameter $\lambda$ in transformers is set to be $0.1$.
The whole model is trained in an end-to-end manner.
The Adam optimizer is chosen to update the model's parameter during the training process.
The hyper-parameter of the Adam optimizer, including learning rate, weight decay, and momentum rates,  are set to be 0.001, 0.01, and (0.9,0.99), respectively.

\begin{figure}[t]
\includegraphics[width=\textwidth]{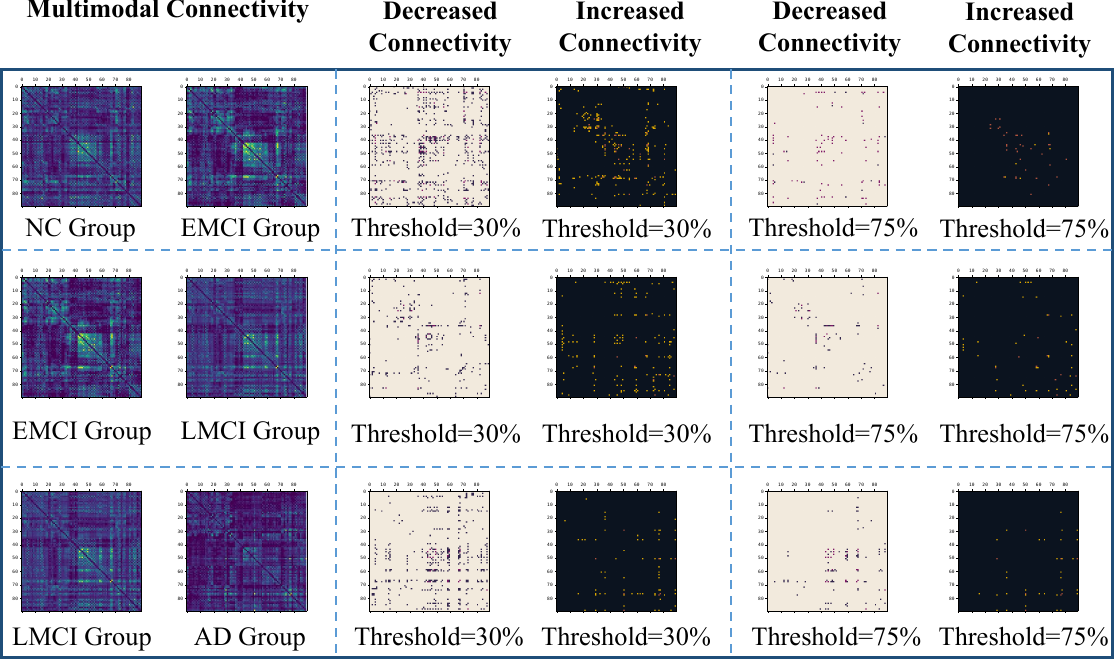}
\caption{Averaged multimodal connectivity of different groups with the same disease stage and the changes of such connectivity under the proceeding of AD stages.
} \label{res}
\end{figure}

\begin{figure}[h]
\includegraphics[width=\textwidth]{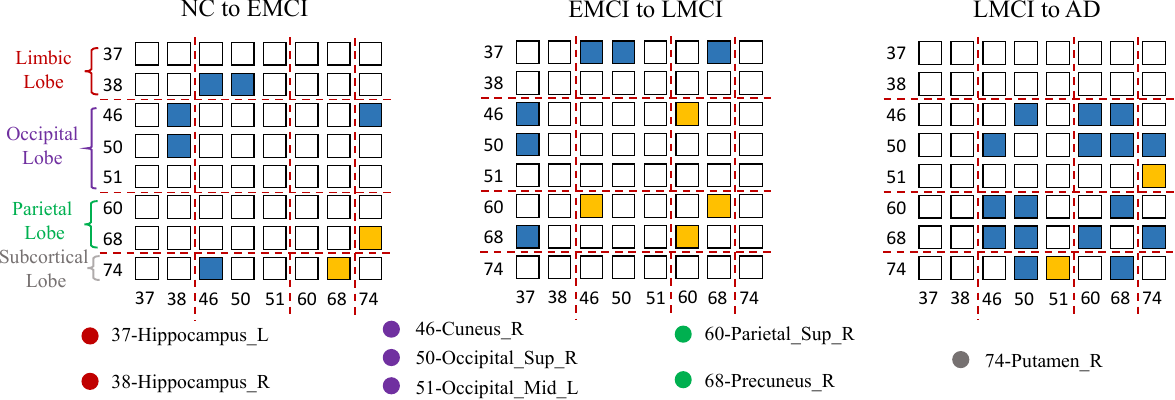}
\caption{The change of multimodal connectivity between the top 8 AD-related ROIs.
The number before ROI is the corresponding AAL id for this ROI.
The blue represents decreased connectivity; The yellow represents increased connectivity.
The red dotted lines divide the 8 ROIs into their corresponding brain lobe.
} \label{res2}
\end{figure}

After applying the trained model to groups of subjects with the same disease stage,
the averaged multimodal connectivity matrices of different disease stages can be obtained.
One of the most important biomarkers of AD is crucial brain connectomes which influence the proceeding of AD stages.
We characterize such connectomes by analyzing the increase/decrease connectivity of the averaged multimodal connectivity matrices of different disease stages.
The visualization of averaged multimodal connectivity matrices and the increase/decrease connectivity with different thresholds are illustrated in Fig.~\ref{res}.
Based on multimodal connectivity matrices, the top 8 ROIs that have the most significant connectivity changing under the AD development process are
37-Hippocampus\_L, 38-Hippocampus\_R, 46-Cuneus\_R, 50-Occipital\_Sup\_R, 51-Occipital\_Mid\_L, 60-Parietal
\_Sup\_R, 68-Precuneus\_R, and 74-Putamen\_R,
where the number before ROI is the corresponding AAL id.
The connectivity changing between the 8 ROIs above is shown in Fig.~\ref{res2}.

To compare the ability to represent AD-related features with different fMRI-DTI fusion models, three binary classification experiments, including
AD vs. NC, LMCI vs. NC, and EMCI vs. NC, are designed.
Three index, detection accuracy(ACC), sensitivity(SEN), and specificity(SPEC), are used to evaluate the models performance.
The experiments' result are shown in Table~\ref{cls}.
\begin{table}[h]\small
\caption{Prediction performance in AD vs. NC, LMCI vs. NC, and EMCI vs. NC. under different fMRI-DTI fusion models}\label{cls}
\resizebox{\textwidth}{20mm}{
\begin{tabular}{|c|c|l|l|l|l|l|l|l|l|l|}
	\hline
	\textbf{Study} \quad & \textbf{Method}  \quad & \multicolumn{3}{|c|}{\textbf{AD vs. NC}}  & \multicolumn{3}{|c|}{\textbf{LMCI vs. NC}} & \multicolumn{3}{|c|}{\textbf{EMCI vs. NC}} \\ \cline{3-11}
    \quad & \quad & \small{ACC} \; & SEN \; & SPE \;& ACC \;& SEN \;& SPE \;& ACC \;& SEN \;& SPE  \;\\
    \hline
    \hline
    Aderghal et al. ~\cite{cls1} & \makecell[c]{CNN+\\ Transfer Learning} &\makecell[c]{\textbf{94.44\%}} &\makecell[c]{\textbf{93.33\%}} &\makecell[c]{\textbf{95.24\%}}&\makecell[c]{87.1\%} &\makecell[c]{87.5\%} &\makecell[c]{86.96\%}&\makecell[c]{85.37\%} &\makecell[c]{88.89\%} &\makecell[c]{82.61\%}
 \\
    \hline
    Dyrba et al.~\cite{cls2}& \makecell[c]{SVM+\\Multi-kernel} &\makecell[c]{86.11\%} &\makecell[c]{85.71\%} &\makecell[c]{86.36\%}&\makecell[c]{83.87\%} &\makecell[c]{72.73\%} &\makecell[c]{90.0\%}&\makecell[c]{82.93\%} &\makecell[c]{84.21\%} &\makecell[c]{81.82\%} \\
    \hline
    Lu et al. ~\cite{cls3}& \makecell[c]{GCN+\\ Deep learning} &\makecell[c]{91.67\%} &\makecell[c]{87.5\%} &\makecell[c]{95.0\%}&\makecell[c]{90.32\%} &\makecell[c]{88.89\%} &\makecell[c]{90.91\%}&\makecell[c]{90.24\%} &\makecell[c]{90.0\%} &\makecell[c]{90.48\%} \\
    \hline
    Proposed & \makecell[c]{GAN+\\ Bi-attention} &\makecell[c]{\textbf{94.44\%}} &\makecell[c]{\textbf{93.33\%}} &\makecell[c]{\textbf{95.24\%}}&\makecell[c]{\textbf{93.55\%}} &\makecell[c]{\textbf{90.0\%}} &\makecell[c]{\textbf{95.24\%}}&\makecell[c]{\textbf{92.68\%}} &\makecell[c]{\textbf{90.48\%}} &\makecell[c]{\textbf{95.0\%}} \\
    \hline
\end{tabular}}
\end{table}
The experiments' result shows that the proposed multimodal fusion model has the advantage of higher accuracy for predicting AD stages than other existing multimodal fusion models.

\section{Conclusion}
In this paper, we proposed a novel CT-GAN to fuse rs-fMRI and DTI to multimodal connectivity of brain network.
The key idea of this work is that we use a bi-attention mechanism to achieve the goal of mutual conversion between structural and functional information.
Therefore, the bi-attention mechanism above can help the proposed model efficiently extracts the complementary information between rs-fMRI and DTI.
The experiments' results proved our multimodal connectivity has higher accuracy of AD prediction than other multimodal fusion methods.
Through analyzing our multimodal connectivity matrices, some AD-related connectomes are obtained. These connectomes are highly consistent with the results of clinical AD studies.
Although this work focuses only on AD,
it is worth mentioning that the proposed model can be easily extended to apply to other neurodegenerative diseases.

\section{Acknowledgement}
This work was supported by the National Natural Science Foundations of China under Grants 62172403 and 81901834, and Shenzhen Key Basic Research Projects under Grant JCYJ20200109115641762.

%
%
%
%

\end{document}